\documentclass[12pt]{article}
\usepackage{graphicx}
\usepackage[top=2cm, bottom=3cm, left=3cm, right=3cm]{geometry}

\title{Puzzles of the Cosmic Ray Anisotropy}
\author{$^{*}$A.D. Erlykin $^{1}$, S.K. Machavariani $^{1}$ and A.W. Wolfendale $^{2}$\\
$(1)$ P N Lebedev Physical Institute, Moscow 119991, Russia.\\
$(2)$ Physics Department, Durham University, Durham, DH1 3LE, UK\\}

\begin{document}
\maketitle

\footnote{$^{*}$Corresponding author: tel +74991358737 \\ 
 E-mail address: erlykin@sci.lebedev.ru}

\begin{abstract}
We discuss three of the known puzzles of the cosmic ray anisotropy in the PeV and sub-PeV energy region. They are 1) the so called inverse anisotropy, 
2) the irregularity in the energy dependence of the amplitude and phase of the first harmonic and 3) the contribution of the single source.
\end{abstract}

\vspace{1mm}


\section{Introduction}
One of the most important and simultaneously most difficult studies of the origin of cosmic rays (CR) is the study of their anisotropy. The difficulty is 
due to the extremely low level of the anisotropy and the steeply falling energy spectrum of CR. In their seminal book 'Origin of 
Cosmic Rays' issued in 1964 V.L.Ginzburg and S.I.Syrovatskii could only give an upper limit of the anisotropy as $<1\%$ \cite{Ginz1}. More than 20 years 
later V.L.Ginzburg and his co-authors on the basis of data obtained by Linsley J. could show that the amplitude and phase of the anisotropy are not 
constant, but vary with the energy \cite{Ginz2}. This variability is the consequence of the non-uniformity of the spatial distribution of CR sources and 
properties of the interstellar medium (ISM). Sources of different locations and ages contribute in different energy regions and magnetic fields of 
different strength and  orientations tend to isotropise the arrival directions of CR particles. 

One of the most interesting regions of the CR energy spectrum is sub-PeV and PeV region where the well-known 'knee' is observed. The study of this region 
is difficult because of the very small intensity of particles and low anisotropy level. That is why there are many unsolved problems and puzzles here 
which still wait for their solution. Below we describe only three of them.
\section{Puzzle 1: an inverse anisotropy}
Due to CR scattering on the chaotic magnetic fields their motion in the Galaxy is like a diffusion from regions of higher to lower CR density. Since the 
solar system is located at about 8 kpc from the galactic center then the main source of CR - supernova remnants (SNR) and pulsars are concentrated in the 
Inner Galaxy. The CR density is higher towards the Galactic Center and their gradient is directed from the Inner to Outer Galaxy. Hence we have to expect 
some anisotropy with the maximum flux from the Inner Galaxy.

\begin{center}
\begin{figure}[hpbt]
\includegraphics[height=15cm,width=14cm,angle=-90]{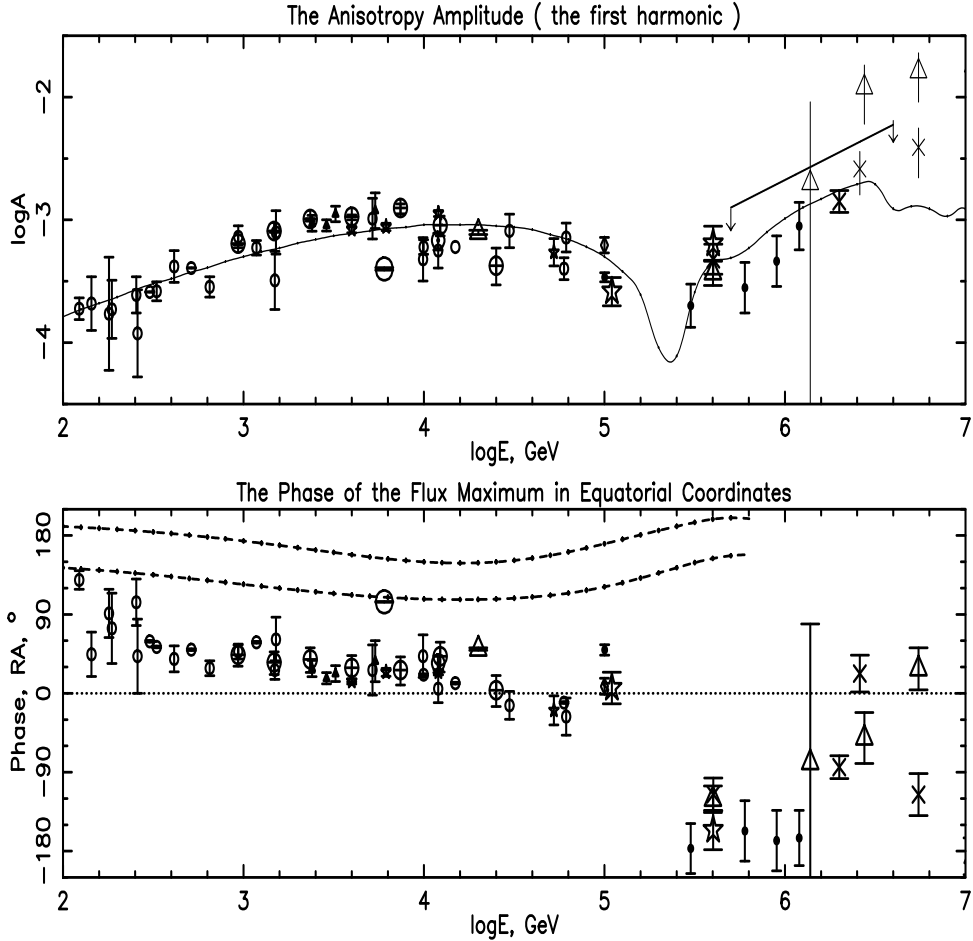}
\caption{\footnotesize The observed amplitude (upper panel) and equatorial phase (lower panel) of the 
first harmonic of the CR anisotropy. Data denoted as $\bigcirc$ are taken from the 
survey presented in \cite{Guil}, $\odot$ - Super-Kamiokande-I \cite{Guil}, $\star$ - Tibet III 
\cite{Amen}, $\triangle$ - Baksan \cite{Kozy}, $\diamond$ - Andyrchi \cite{Kozy},
$\bullet$ - EAS-TOP \cite{Agli}. Thick line above $logE = 5.7$ with an arrow - upper 
limits of the amplitude, given by KASCADE \cite{Anto}. Here and in the text below the energy $E$ is in GeV. The full line in the upper panel relates to 
calculations with our model described in Section 3 \cite{EW1}.  The area between thin dashed lines
 in the lower panel denotes the region in {\em galactic} coordinates corresponding to the RA region occupied
by experimental points in equatorial coordinates assuming that the observations were made at declinations about 30$^\circ$-60$^\circ$.} 
\label{fig:fig01}
\end{figure}
\end{center}

The expected and observed situation with the first harmonic of the anisotropy in the sub-TeV region is shown in Figure 1.  
Ihe upper panel shows the amplitudes and it is seen that they do not exceed the value of $10^{-3}$. The lower panel presents measurements of the phase 
in equatorial coordinates. If transferred to galactic coordinates they occupy the region delimited by two dashed lines. It is seen that when the expected 
phase of the maximum CR flux is from the Inner Galaxy with $\ell \approx 0^\circ$ \cite{EW1}, the observed phase occupies the longitude $\ell$-region 
between $90^\circ$ and $-90^\circ$, i.e. CR come preferentially from the Outer Galaxy. We call this phenomenon an 'inverse anisotropy'.           

One of the possible explanations of an inverse anisotropy is to assume that it is a local phenomenon caused by a spatial orientation of the magnetic 
field or the ISM density fluctuation. The Sun is located in the Local Bubble at the inner edge of the Orion spiral arm. The strength of the magnetic 
field and of the ISM density in the arm is higher than in the interarm region where the Local Bubble is located. The diffusion in the nearby Outer side 
of the Galaxy is slower than locally. CR moving from the Inner Galaxy meet like a wall, a part of them reflect from it, accumulate in a number and create 
an inverse gradient \cite{EW2}.

There might be an alternative explanation \cite{Ahlers}. Recent measurements of CR at sub-TeV - sub-PeV energies demonstrated the irregular non-power law 
shape of the energy spectrum. As an example the spectral hardening at ~200-300 GeVnoticed in CREAM, PAMELA, AMS-2 experiments can be mentioned, the 
steepening at the magnetic rigidity of ~10TV in the NUCLEON experiment and others. The observed irregularities point to the possible role of local 
sources. However, to be responsible for an observed inverse anisotropy these local sources should be located mostly in the Outer Galaxy. 

 We should underline that these explanations are only a possible assumption with many internal uncertainties. However, we should emphasize that the 
inverse anisotropy is most likely a local phenomenon, which is caused by the reflection of CR from a nearby region of higher ISM density or the dominance 
of some local sources in the Outer Galaxy.     
\section{Puzzle 2: peculiarity of the amplitude and phase}
In general there are several features noticable in Figure 1: \\
(i) a good consistency of the results at energies up to a few PeV; \\
(ii) the extremely small $\sim 10^{-4} \div 10^{-3}$ amplitude of the anisotropy; \\
(iii) the visible rise of the amplitude $A$ with energy $E$ up to $logE \approx 4$; \\
(iv) a moderate fall of the amplitude above $logE \approx 4$ up to a minimum at 
$logE \approx 5.3 \div 5.5$; \\
(v) the rise of the amplitude beyond this minimum up to a few PeV; \\ 
(vi) the approximately constant phase at low energies which suddenly changes its 
direction at about the same energy of $logE \approx 5.3 \div 5.5$ 
where the amplitude has a minimum; \\
(vii) in the PeV region, where the rise of the amplitude is observed, the phase has 
an apparent trend to recover up to its previous direction close to $RA \sim 0$.   

In what follows we shall endeavour to build a model which can reproduce these features 
with the minimum number of assumptions. This model contains three basic ingredients:
the Galactic Disk, the Halo and the Single Source (SS). Although we separate here the 
role of the Single Source, we understand that, in fact, it is just part of CR in 
the Disk.

The details of the model suggested to explain all these peculiar features are given in 
\cite{EW1}. Here we give just its main features and they are illustrated in Figure 2.
\begin{figure}[ht]
\begin{center}
\includegraphics[height=15cm,width=12cm,angle=-90]{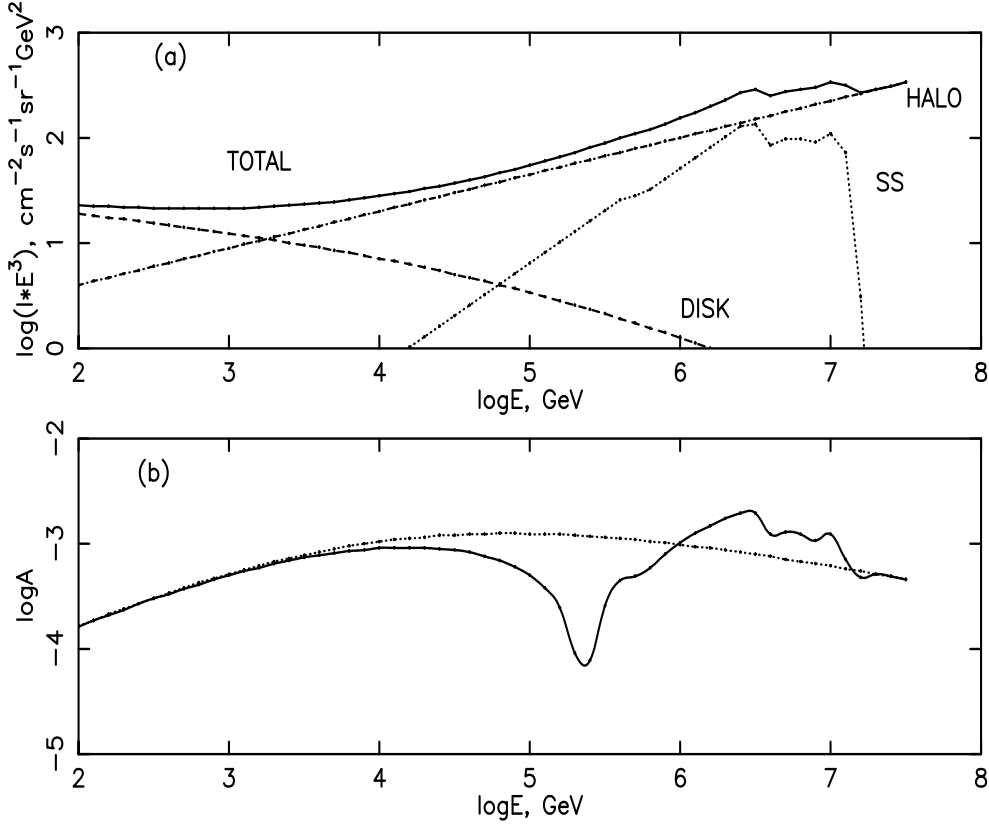}
\end{center}
\caption{\footnotesize (a) Schematic formation of the CR energy spectrum from the Disk,
 the Halo and the Single Source. 
(b) The amplitude of the first harmonic of the CR anisotropy obtained with 
contributions from Disk + Halo + Single Source (~full line~) and the same with 
contributions only from Disk + Halo without Single Source (~dotted line~).}   
\label{fig:fig2}
\end{figure}   
We assume that the anisotropy appears only in the vicinity of the sources, i.e. in the
 Disk. The amplitude of the anisotropy $A$ is connected with the CR intensity $I$, its 
gradient, $gradI$, and the diffusion coefficient $D$ as $A=\frac{3DgradI}{cI}$. If the 
relative gradient $\frac{gradI}{I} \neq 0$ and $D \propto E^{\delta_d}$ then $A$ in the
 Disk rises with energy as $A_d \propto E^{\delta_d}$. The anisotropy of CR in the Halo
 is postulated as being $A_h = 0$ and the isotropy of CR re-entrant from the Halo back 
into the Disk dilutes the anisotropy of CR produced and trapped in the Disk.

In this treatment we just consider the first harmonic. Later work will deal with higher
 multipoles. We calculate the amplitude of the first harmonic for the case where only 
Disk and Halo contribute to CR as 
\begin{equation}
A = \frac{A_dI_d}{I_d+I_h}
\end{equation}
The result is shown in Figure 2b by the dotted line. The rise of the amplitude at 
energies above 100 GeV is due to the rise of the diffusion coefficient in the 
expression for $A_d$ mentioned above. The slow decrease of $A$ above $\sim$10 TeV is 
explained by the rising fraction of isotropic CR from the Halo, which overcomes the rise of $A_d$.
However, this scenario does not reproduce the remarkable dip in the amplitude visible 
in the experimental data at $logE = 5.3-5.5$ and the subsequent rise of the 
amplitude above this dip (~Figure 1a~).

We think that these features, if they are real, are connected with the existence 
of the Single Source, from which the CR energy spectrum is schematically shown by the 
dotted line in Figure 2a and denoted as 'SS'. However, this idea alone is not enough to
 reproduce the experimental data and here the examination of the phase of the first 
harmonic could help. In Figure 2b it is seen that after the moderate decrease in the 
energy interval 0.1 - 100 TeV the phase suddenly changes to its opposite. We consider 
this change seriously and propose that CR from the Single Source have a phase opposite 
to that of the background at lower energies. This is a necessary complementary 
requirement in our model. The rising part of CR coming from the opposite direction 
would reduce the anisotropy of the background from the Disk and Halo as
\begin{equation}
A = |\frac{A_dI_d-A_{ss}I_{ss}}{I_d+I_h+I_{ss}}|
\end{equation}
The result of the calculations with the contribution of the Single Source is shown in 
Figure 2b by the full line. A comparison with the experimental data is shown also in 
Figure 1a by the full line. It is seen that after minimum in the dip the amplitude 
of the anisotropy starts rising again and it is caused by the rising contribution of 
the Single Source, which has the opposite phase.
 
We think that the described model with three basic ingredients: Disk, Halo and Single 
Source is reasonable and is worthy of discussion. The new features advocated here are:

(a) The dominance of the Halo component in the sub-PeV region. It means that the CR 
which we observe and study in spite of being ourselves inside the Disk come mostly from
 the Halo. It is disputable but helps to understand the low anisotropy, small radial 
gradient of CR intensity and small level of irregularities in the regular power law 
energy spectrum. 

(b) The idea about the Single Source, which has to be nearby and young and creates the 
knee, usually raises questions: 'if it is nearby why do'nt we see it in the 
anisotropy ?'. It is a very reasonable question and this work gives the answer. The 
Single Source causes the stronger decrease and the dip in the amplitude of the dipole 
anisotropy
at sub-PeV energies. It is also seen in the change of the phase of the anisotropy at 
the same energies. It means that the Single Source should deliver CR from the direction
opposite to the direction of CR from the background and it is a new assumption in the 
Single Source scenario. Above the dip energy the amplitude starts to rise again with 
the opposite phase, as expected.

The dip in this model appears as the result of subtraction of two 
bigger values. Its position and shape are extremely sensitive to the choice of
parameters participating in expression (2). The relatively good agreement with the 
experimental data is the result of the fitting procedure, but, nevertheless: \\
(i) it demonstrates the possibility of achieving agreement within the 
framework of our simplistic model and \\
(ii) the high sensitivity of the dip to the input parameters of the expression (2) 
gives the possibility of investigating these parameters when precise results in this 
energy region are obtained.         

(iii) The phase of the first harmonic in the PeV region, where the contribution of the
Single Source is big enough, could help to locate it on the sky. The present 
experimental data have a too big spread to make a conclusion.  
   
We understand that this scenario raises more questions than gives answers. For 
instance, the main questions are:

(a) Do the Halo and the Single Source really exist ? Arguments for positive answers
are given in the \S3, but more supportive arguments are needed.

(b) Why numerical estimates for CR intensity in the Disk and the Halo, calculated
with the expression (2), are smaller than in the observations~?
    
(c) Why is the energy spectrum of CR in the Disk steeper than the spectrum in the Halo 
~? According to our conception developed in \cite{EW1} the spectrum in the Disk, with 
its higher turbulence in the interstellar medium due to SN explosions, should be flatter
 than in the Halo where there are no such powerful sources of turbulence as SN.

(d) To what extent are the simplified assumptions about the shapes and normalisation of
 the Disk, Halo and Single Source spectra as well as other parameters: $A_d, A_h$ and 
$A_{ss}$, reasonable and what will the more sophisticated approach do for the result ?

The answers to these and other puzzling questions are the subject of further work.
\section{Puzzle 3: the nature of the Single Source}
The existence of the Single Source has been proposed by us to explain the puzzling 
sharpness of the knee in the size spectrum of extensive air showers (EAS) ( see \cite{EW4} and later publications 
). The physical basis of this proposal is the evident non-uniformity of the spatial and
 temporal distributions of SN explosions and subsequent SNR. As a result one SN could 
explode not very long ago and close to the solar system.. Its contribution to the CR 
intensity is rather high and it gives rise to a small peak (~knee~) above the 
background from other SNR - it is our Single Source. 

To search for the Single Source authors \cite{Pavl} used the so called difference metod.
It is different from the traditional study of the CR intensity.
Its main idea is that the difference between properties of EAS coming from the direction 
of the Single Source and from the opposite direction should be maximum. The method is stable against random 
experimental errors and allows to separate anomalies connected with the laboratory coordinate system from anomalies
in the celestial coordinate system. The method allows to study the whole celestial sphere including regions outside 
of the line of sight of the experimental installation.

To search for an anisotropy authors \cite{Pavl} used experimental data obtained with the GAMMA experimental EAS array and have taken
the EAS age parameter as a characteristic of the EAS property. The difference between EAS age distributions in two 
opposite directions was quantified by the reduced $\chi^2$ value \cite{Pavl}. The result of the search is given in 
Figure 3.    
\begin{figure}[ht]
\begin{center}
\includegraphics[height=10cm,width=15cm]{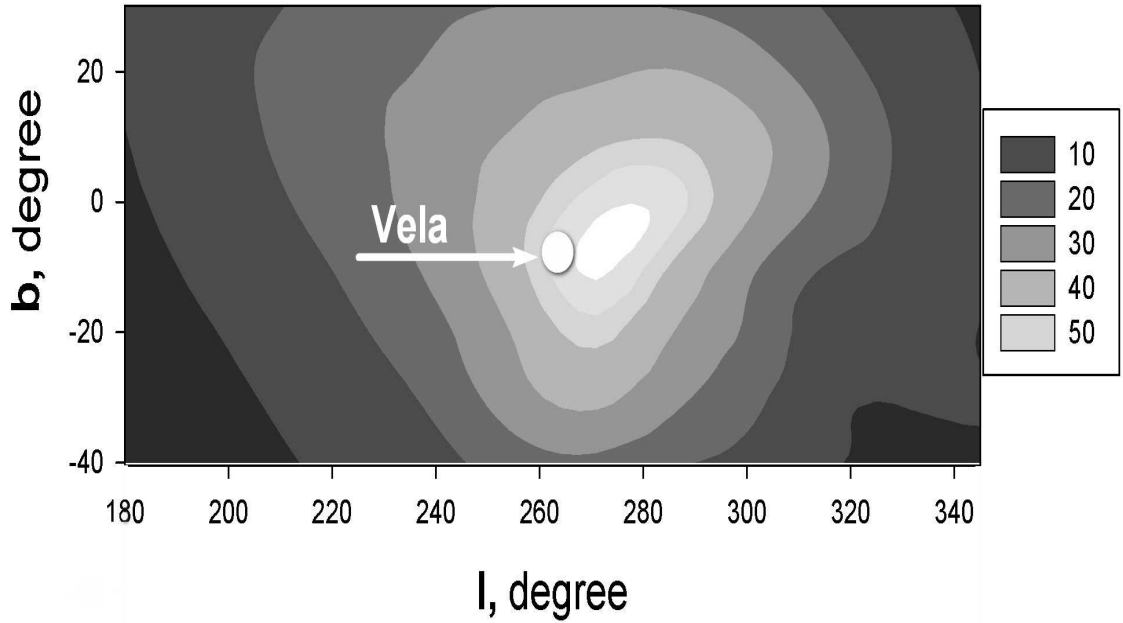}
\end{center}
\caption{\footnotesize Two-dimensional profile of the reduced $\chi^2$ of the difference between age distributions of EAS coming from opposite directions 
in galactic coordinates.}   
\label{fig:fig3}
\end{figure}   
The maximum difference in galactic coordinates has been found at $\ell=277\pm3^\circ, b=-5\pm3^\circ$. The closest source to this location is cluster 
Vela and therefore it is a good candidate for the role of the Single Source, which is responsible for the knee in the CR spectrum and the minimum 
in the amplitude of the anisotropy. However, a firm conclusion can be drawn only if it will be confirmed by the 
analysis of data from other EAS arrays.

An alternative explanation is the influence of the regular magnetic field in the area surrounding the Earth especially taking into account the nearby 
spiral arm (~see Puzzle 1 of this paper~). The only point we should like to stress is that the effect of the regular magnetic field in PeV and sub-PeV 
energies should be small compared with the effect of the single source in order not to destroy the diffusive character of CR propagation. Additional 
asguments in favour of the dominance of the single source in the formation of the maximum difference in age distributions of EAS coming from the opposite 
directions could be the vicinity of Vela source and the 'younger' age of EAS coming from the direction of the maximum since magnetic fields do not change 
the energy spectrum of EAS and their ages.

In the section 6 of our paper \cite{Pavl} we speculated that a minor shift of the maximum difference from Vela source seen in Figure 3 could be due to the
influence of this small regular magnetic field.  
\section{Conclusion}
We mentioned only 3 puzzles existing in the PeV and sub-PeV energy region. However, there are many other there and after Prince Hamlet we can conclude
that 'there are many things in heaven and earth, Horatio, than are dreamt of in your philosophy'.

\end{document}